\documentclass{epl}

\usepackage{graphics}
\usepackage{graphicx}
\usepackage{amsmath}
\usepackage{psfrag}
\usepackage{subfigure}
\usepackage{amsfonts}
\newcommand{\dd}{\text{d}}

\newcommand{\pop}[2]{\frac{\partial #1}{\partial #2}}
\newcommand{\ket}[1]{|#1\rangle} 
\newcommand{\bra}[1]{\langle#1|}

\newcommand{\erwwert}[1]{\langle#1\rangle}

\title{On the concept of pressure in quantum mechanics}
\shorttitle{Pressure in quantum mechanics}
\author{P. Borowski\thanks{E-mail: \email{borowski@theo1.physik.uni-stuttgart.de}} \and J. Gemmer \and G. Mahler}
\shortauthor{P. Borowski \etal}
\institute{Institut f\"ur Theoretische Physik I - Universit\"at Stuttgart, Pfaffenwaldring 57, 70550 Stuttgart, Germany}
\pacs{03.65.Ge}{Solutions of wave equations: bound states}
\pacs{03.65.Ta}{Foundations of quantum mechanics; measurement theory}

\begin{document}

\maketitle

\begin{abstract}
Heat and work are fundamental concepts for thermodynamical systems. When these are scaled down to the quantum level they require appropriate embeddings. Here we show that the dependence of the particle spectrum on system size giving rise to a formal definition of pressure can, indeed, be correlated with an external mechanical degree of freedom, modelled as a spatial coordinate of a quantum oscillator. Under specific conditions this correlation is reminiscent of that occurring in the classical manometer.
\end{abstract}


\section{Introduction}
\label{intro}
Gibbs fundamental relation~\cite{Roemer},
\begin{equation}
\label{eq_dE}
\dd E = T\,\dd S + \sum_j \xi_j \,\dd X_j\,,
\end{equation}
describes the change of the internal energy $E$ of a thermodynamic system in response to a change of its entropy $S$ and of the general extensive working variables $X_j$. Temperature $T$ and the $\xi_j$ are the corresponding conjugated intensive variables. This relation can be given a statistical interpretation.

In statistical mechanics, thermodynamic values like the energy are interpreted as the mean over all micro-state values $E_i$ realized with the probability $w_i$:
\begin{equation}
\label{eq_overline_E}
\overline{E}=\sum_i w_i E_i\,.
\end{equation}
Based on $w_i$ the entropy is defined as
\begin{equation}
S=-k_B\sum_iw_i\ln w_i\,.
\end{equation}
In thermodynamic equilibrium, the energy distribution of a canonical system follows a Boltzmann-distribution:
\begin{equation}
\label{eq_Boltzmann}
w_i=\frac{\exp(-\beta E_i)}{\sum_j \exp(-\beta E_j)}
\end{equation}
with $\beta = \frac{1}{k_B T}$, $k_B$ being the Boltzmann-constant and $T$ the temperature.

Finally, assuming that the single energy levels $E_i$ depend on a set of parameters $\{X_j\}$, we recover eq.~(\ref{eq_dE}) from eq.~(\ref{eq_overline_E}):
\begin{equation}
\label{eq_overline_E_2}
\dd \overline{E} = -\frac{1}{\beta}\sum_i \,\dd w_i \ln w_i + \sum_j \xi_j\,\dd X_j\,.
\end{equation}
A somewhat similar approach for classical Hamiltonian systems can be found in~\cite{Rugh}.

Eq.(\ref{eq_overline_E_2}) tells us that there exist two fundamentally different ways of changing the mean energy in a thermodynamic system:
\begin{enumerate}
\item Changing the state occupation $w_i(E_i)$. This changes the entropy of the system, which, for large, thermodynamic systems, corresponds to a change of temperature (quasi-stationary changes are assumed to maintain the Boltzmann-character (eq.~(\ref{eq_Boltzmann})) of the distribution, cf.~\cite{Opatrny}).
\item Changing the energy levels $E_i(\{X_j\})$ by means of the parameters $X_j$. For $w_i = \text{const.}$ this amounts to an adiabatic change in terms of work.
\end{enumerate}
However, in the quantum regime, both these processes have to be reconsidered because it is not clear from the beginning, what one should call intensive parameters ($T$, $X_j$) or whether thermodynamic relations derived for classical systems can still be taken over into the quantum domain.

For quantum systems the first process requires an explicit thermodynamic embedding, as has been studied recently~\cite{Jochen1}\cite{Jochen2}\cite{ich}. The second process will be shown to require what could be called a mechanical embedding.

In this paper a simple two-particle quantum system will be used to investigate the dependence of the spectrum of one of the subsystems on a spatial coordinate of the other subsystem that could be considered as an external parameter $X_j$. This will be done from first principles using stationary perturbation theory only. For specific choices of the two perturbation parameters introduced we will find not only a quantum analog to the classical definition of pressure $p$ in thermodynamics, $p=-\pop{E}{V}$, but also its correlation with the mechanical embedding.

Harmonically bound mirrors have also been studied in the context of measurement-scenarios for radiation- and gravitational- field pressure. In these cases, though, the concept of pressure has not been challenged~\cite{Fabre}\cite{Pace}\cite{Vitali}.


\section{The model: A quantum-manometer}
\label{sec:1}
\begin{figure}
\psfrag{PiB}[cb]{$g$}
\psfrag{HO}{W}
\psfrag{x_1,M_1}{$x_g,M_g$}
\psfrag{x_W,M_W}{$x_{\text{\tiny W}},M_{\text{\tiny W}}$}
\psfrag{V}{$V$}
\psfrag{L+x_W}{$L+x_{\text{{\tiny W}}}$}
\psfrag{f}{$f_{\text{W}}$}
\psfrag{0}{0}
\psfrag{x}{$x$}
\onefigure[width = 5cm]{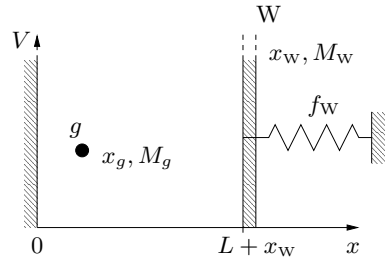}
\caption{A one-dimensional 'quantum-manometer' with the two subsystems 'gas' $g$ and 'wall' W.}
\label{fig_manometer}
\end{figure}

We are going to analyze some aspects of a quantum version of a classical pressure-registering device, the manometer (fig. \ref{fig_manometer}).
The total model will be taken as a two-particle system, consisting of a 1-particle-'gas', $g$, with particle mass $M_g$ at the position $x_g$ in an infinitely high square well potential of linear dimension $L$, and a piston, i.e. a wall, W, of mass $M_{\text{W}}$ and position $x_{\text{W}}$ that forms one of the two boundaries of the square well. The wall is connected to a spring with the spring constant $f_{\text{W}}$ and therefore moving in a harmonic potential. We will be looking for correlations between the energy states of the gas and the average position of the piston. Note that this collective degree of freedom of the wall is but one part of the dynamical reaction of the container walls. The other microscopic degrees of freedom would give rise to a thermodynamical embedding~\cite{Jochen1}\cite{Jochen2}\cite{ich}\cite{Jochen_Diplomarbeit}.

If the system was uncoupled and the wall was not allowed to move ($x_{\text{W}}=0$), the Hamiltonian of the system would be just the sum of a particle-in-a-box ($g$) Hamiltonian with quantum number $j_g$ and the Hamiltonian of an harmonic oscillator (the reacting wall W) with quantum number $j_{\text{W}}$. The total energy would then be $E^{(0)}_{j_g,j_{\text{W}}}=E^{g}_{j_g}+E^{\text{W}}_{j_{\text{W}}}=\frac{\pi^2\hbar^2j_g^2}{2M_gL^2}+\hbar\sqrt{\frac{f_{\text{W}}}{M_{\text{W}}}}\left(j_{\text{W}}+\frac{1}{2}\right)$. Taking the position of the wall as a dynamical variable of the total system, the two subsystems are coupled, and the new state of the total system may be described in terms of stationary perturbation theory. The perturbed system eigenstates will be \cite{Jochen_Diplomarbeit}
\begin{equation}
\label{gl_QM_Störungsrechnung}
\ket{j_g,j_{\text{W}}}' \approx \ket{j_g,j_{\text{W}}}^0 + \sum_{\substack{k_g,k_{\text{W}} \\ k_g\ne j_g \, \vee \, k_{\text{W}}\ne j_{\text{W}}}}{}^1C^{j_gj_{\text{W}}}_{k_gk_{\text{W}}}\ket{k_g,k_{\text{W}}}^0
\end{equation}
with the first order perturbation coefficients
\begin{equation}
\label{gl_QM_C}
{}^1C^{j_gj_{\text{W}}}_{k_gk_{\text{W}}} = \frac{{}^0\bra{k_g,k_{\text{W}}}\hat{W}\ket{j_g,j_{\text{W}}}^0}{E^{(0)}_{j_g,j_{\text{W}}}-E^{(0)}_{k_g,k_{\text{W}}}} = \frac{{}^0\bra{k_g,k_{\text{W}}}\hat{W}\ket{j_g,j_{\text{W}}}^0}{E^{g}_{j_g}+E^{\text{W}}_{j_{\text{W}}}-E^{g}_{k_g}-E^{\text{W}}_{k_{\text{W}}}}.
\end{equation}

The perturbation operator $\hat{W}$ is the difference between the Hamiltonians for the uncoupled ($x_{\text{W}}=0$) and the coupled ($x_{\text{W}}\ne 0$) case and since the wall represents an infinite potential for the particle in the box (gas $g$), $\hat{W}$ is a positive or negative infinite rectangular potential with dimension $|x_{\text{W}}|$ for which perturbation theory as in eq.~(\ref{gl_QM_Störungsrechnung}) does not make sense (the overlap integrals will all be either zero or infinite).

However, by appropriate coordinate transformation it is possible to shift the perturbation from the potential terms in the Hamiltonian to the kinetic part. The transformation
\begin{figure}
\psfrag{xlabem}[]{$x_g$}
\psfrag{ylabem}[]{$x_{\text{W}}$}
\psfrag{zlabem}[b]{$V$}
\psfrag{0}[b]{0}
\psfrag{l1}[l]{$L$}
\psfrag{l2}[r]{$-L$}
\psfrag{xlabel}[]{$y_g$}
\psfrag{ylabel}[]{$y_{\text{W}}$}
\psfrag{zlabel}[b]{$V$}
\psfrag{0}[b]{0}
\psfrag{l1}[l]{$L$}
\psfrag{l2}[r]{$-L$}
\twofigures[width=4.5cm]{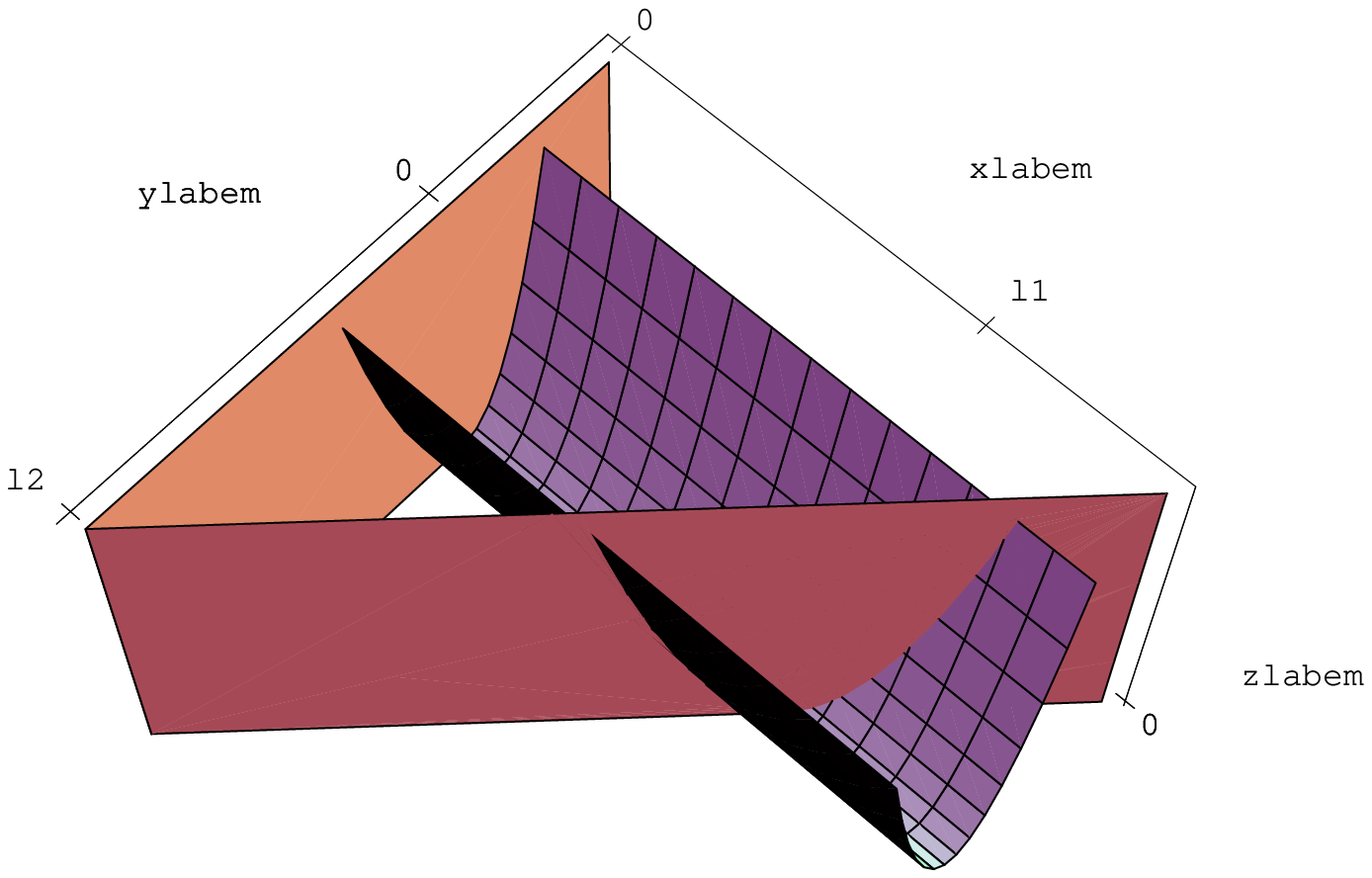}{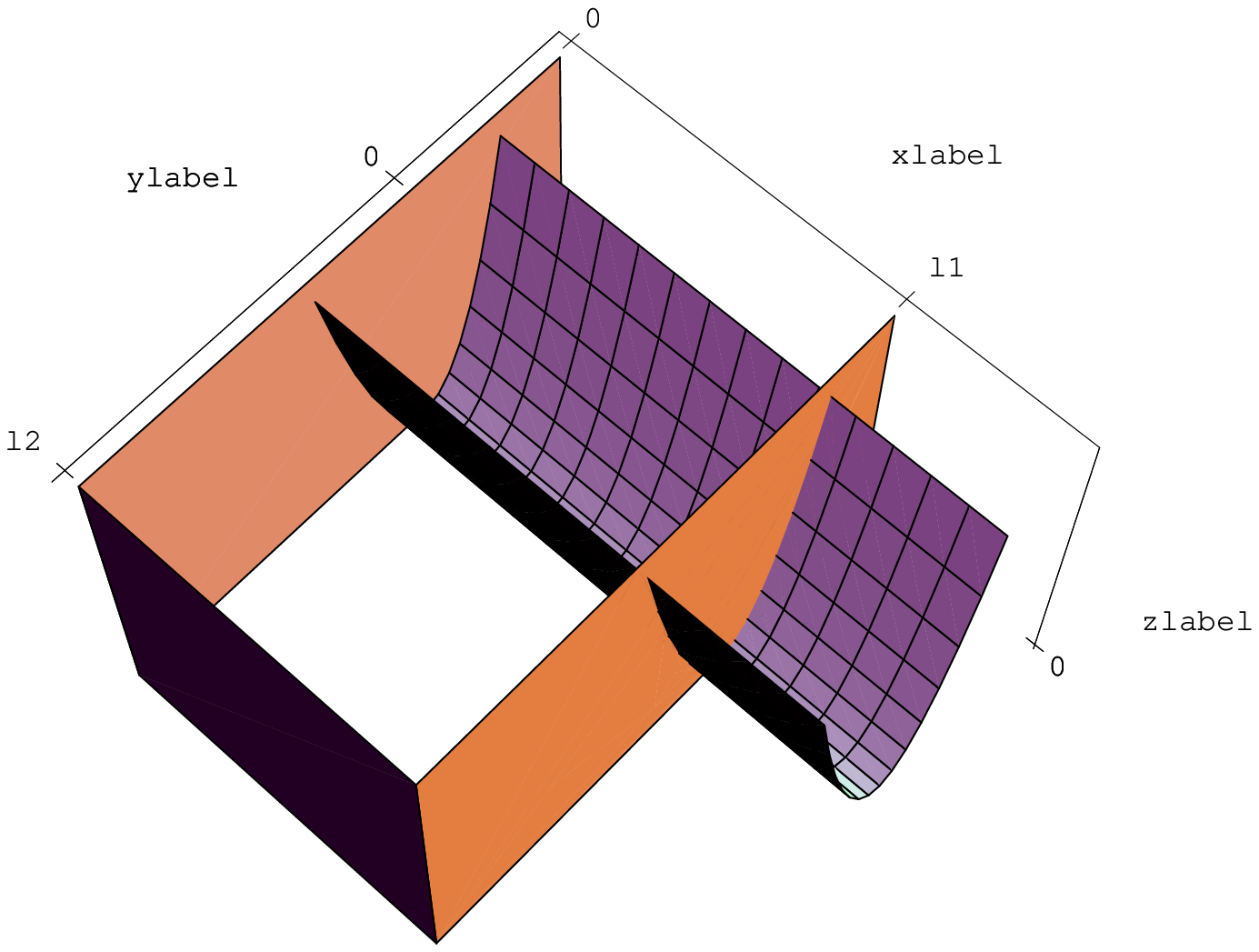}
\caption{Illustration of the coordinate transformation~(\ref{eq_transformation}). The vertical areas represent the boundary of the infinite potential region. Potential in the original $\{x_g,x_{\text{W}}\}$ coordinate system.}
\label{fig_pot_x}
\caption{The Potential after the transformation in the $\{y_g,y_{\text{W}}\}$ coordinate system.}
\label{fig_pot_y}
\end{figure}
\begin{equation}
y_g = \frac{x_g}{x_{\text{W}}+L}L, \quad y_{\text{W}} = x_{\text{W}} \quad \text{(illustrated in figs.~\ref{fig_pot_x} and \ref{fig_pot_y})}
\label{eq_transformation}
\end{equation}
together with the condition $x_{\text{W}}\ll L$ (the wall reacts only slightly) leads to a simple uncoupled potential in the transformed Hamiltonian: An infinite square well from $y_g=0$ to $y_g=L$, $V^g(y_g)$, plus a harmonic potential in $y_{\text{W}}$, $V^{\text{W}}(y_{\text{W}})=\frac{1}{2}f_{\text{W}}y_{\text{W}}^2$. Together with the transformation of the two derivatives of second order in the original Hamiltonian, and an expansion of $y_{\text{W}}$ around zero to first order, the new, transformed Hamiltonian can be written as
\begin{align}
\label{gl_QM_H(y)_3}
& \hat{H}(y_g,y_{\text{W}}) \approx \underbrace{-\frac{\hbar^2}{2M_g}\frac{\partial^2}{\partial y_g^2}+ V^{g}(y_g) -\frac{\hbar^2}{2M_{\text{W}}}\frac{\partial^2}{\partial y_{\text{W}}^2}+V^{\text{W}}(y_{\text{W}})}_{\hat{H}^{(0)}} \underbrace{-\frac{\hbar^2}{2M_{\text{W}}}\frac{y_g^2}{L^2}\frac{\partial^2}{\partial y_g^2}}_{\hat{W}_a}\underbrace{-\frac{\hbar^2}{M_{\text{W}}}\frac{y_g}{L^2}\frac{\partial}{\partial y_g}}_{\hat{W}_b} \nonumber \\
& +\underbrace{\frac{\hbar^2}{M_g}\frac{y_{\text{W}}}{L}\frac{\partial^2}{\partial y_g^2}}_{\hat{W}_c} + \underbrace{\frac{\hbar^2}{M_{\text{W}}}\frac{y_g^2y_{\text{W}}}{L^3}\frac{\partial^2}{\partial y_g^2}}_{\hat{W}_d}+\underbrace{\frac{\hbar^2}{M_{\text{W}}}\frac{2y_gy_{\text{W}}}{L^3}\frac{\partial}{\partial y_g}}_{\hat{W}_e} +\underbrace{\frac{\hbar^2}{M_{\text{W}}}\frac{y_g}{L}\frac{\partial^2}{\partial y_g \partial y_{\text{W}}}}_{\hat{W}_f} \underbrace{- \frac{\hbar^2}{M_{\text{W}}}\frac{y_gy_{\text{W}}}{L^2}\frac{\partial^2}{\partial y_g \partial y_{\text{W}}}}_{\hat{W}_h}\,.
\end{align}
$\hat{H}^{(0)}$ is the Hamiltonian for the uncoupled system of a particle in a box and a harmonic oscillator in the $\{y_g,y_{\text{W}}\}$-system. The seven terms $\hat{W}_a, ..., \hat{W}_h$ are the parts of the new perturbation operator $\hat{W}$. Their weights have to be examined in the following by looking at the seven contributions to the perturbation coefficient of first order, ${}^1C^{j_gj_{\text{W}}}_{k_gk_{\text{W}}}$ (eq.~(\ref{gl_QM_C})). From now on, only the case $j_{\text{W}}=0$ will be considered.

After introducing the two dimensionless expansion parameters
\begin{equation}
\label{gl_QM_lambda_beta}
\lambda=\frac{\hbar^{\frac{1}{2}}}{M_{\text{W}}^{\frac{1}{4}}f_{\text{W}}^{\frac{1}{4}}L} \quad \text{and} \quad \beta=\sqrt{\frac{M_g}{M_{\text{W}}}}
\end{equation}
it follows that five of the seven contributions are of order two or three in these parameters and only two, ${}^1_cC^{j_g0}_{k_gk_{\text{W}}}$ and ${}^1_fC^{j_g0}_{k_gk_{\text{W}}}$ (originating from the two contributions $\hat{W}_c$ and $\hat{W}_f$ to $\hat{W}$) are of first order in $\{\lambda,\beta\}$:
\begin{equation}
\label{eq_C_1}
{}^1_cC^{j_g0}_{k_gk_{\text{W}}}=\frac{\pi^2j_g^2}{\sqrt{2}}\frac{\lambda^3}{\beta^2}\delta_{k_{\text{W}}1}\delta_{j_gk_g}
\end{equation}
\begin{equation}
\label{eq_C_2}
{}^1_fC^{j_g0}_{j_gk_{\text{W}}}=-\frac{1}{2\sqrt{2}}\lambda\delta_{k_{\text{W}}1}\,; \qquad {}^1_fC^{j_g0}_{k_gk_{\text{W}}}=-\frac{2\sqrt{2}j_gk_g(-1)^{j_g+k_g}}{\left[\pi^2(j_g^2-k_g^2)-2\frac{\beta^2}{\lambda^2}\right](j_g^2-k_g^2)}\frac{\beta^2}{\lambda}\delta_{k_{\text{W}}1}\,.
\end{equation}
The parameters $\lambda$ and $\beta$ can be made arbitrarily small by choosing $M_{\text{W}}\gg M_g$ and $L$ and $f_{\text{W}}$ large. They should both be of the same order of magnitude to make this perturbation theory work but that is just a sufficient condition and does not mean at all that cases with $\lambda \not \approx \beta$ would not work. The normalization factor
\begin{equation}
\label{eq_normalization_factor}
N=\sqrt{1+\sum_{\substack{k_g,k_{\text{W}} \\ k_g\ne j_g \, \vee \, k_{\text{W}} \ne j_{\text{W}}}}({}^1C^{j_gj_{\text{W}}}_{k_gk_{\text{W}}})^2}
\end{equation}
for the new perturbed states (\ref{gl_QM_Störungsrechnung}) is of second order in $\{\lambda, \beta\}$ and can therefore be taken as one ($N\approx 1$).

As a test for this perturbation theory, the interaction energy of first order,
\begin{equation}
\label{eq_perturbation_energy}
E_{j_g0}^{(1)}={}^0\bra{j_g,0}\hat{W}\ket{j_g,0}^0\,,
\end{equation}
can be calculated and it follows, that (\ref{eq_perturbation_energy}) is equal to zero. The second order perturbation energy divided by the energy of the uncoupled system is, again, of second order in $\{\lambda, \beta\}$.

For calculating the expectation value $\erwwert{x_{\text{W}}}$ of the position of the wall, the functional determinant has to be regarded when transforming the differentials $\dd x_g$ and $\dd x_{\text{W}}$ to $\dd y_g$ and $\dd y_{\text{W}}$ in the integrals. After neglecting terms of second order in ${}^1C^{j_g0}_{k_gk_{\text{W}}}$ it follows
\begin{equation}
\label{eq_erwwert_x}
\erwwert{x_{\text{W}}}_{j_g}\approx {}^{\,\,0}_{\text{W}}\bra{0}\left( \frac{y_{\text{W}}^2}{L}+y_{\text{W}}\right) \left[ \ket{0}^0_{\text{W}}+2\sum_{k_{\text{W}}\ne 0}{}^1C^{j_g0}_{j_gk_{\text{W}}}\ket{k_{\text{W}}}^0_{\text{W}}\right]
\end{equation}
with $\ket{k_{\text{W}}}^0_{\text{W}}$ being the unperturbed eigenstates of a harmonic oscillator in the $y_{\text{W}}$ coordinate. Looking at the relations between those eigenstates and observing (\ref{eq_C_1}) and (\ref{eq_C_2}) we get
\begin{equation}
\label{gl_QM_erwwert_x_W_ergebnis}
\erwwert{x_{\text{W}}}_{j_g}=\frac{\pi^2\hbar^2j_g^2}{M_gL^3f_{\text{W}}}.
\end{equation}
This, however, is exactly what one would expect based on the classical expression $F=-\pop{E(L)}{L}$: The force $F$ would shorten the spring with spring constant $f_{\text{W}}$ by $d=\frac{F}{f_{\text{W}}}$ and for a particle in a box with
\begin{equation}
E^{g}_{j_g}=\frac{\pi^2\hbar^2 j_g^2}{2M_gL^2} \quad \Rightarrow \quad F^{g}_{j_g}=\frac{\pi^2\hbar^2 j_g^2}{M_gL^3}
\end{equation}
this $d$ equals exactly the expectation value of eq. (\ref{gl_QM_erwwert_x_W_ergebnis}):
\begin{equation}
\boxed{F_{j_g}^{g}=-\pop{E_{j_g}^{g}}{L}=f_{\text{W}}\erwwert{x_{\text{W}}}_{j_g}}\,.
\end{equation}
Note that this result is valid also, if the gas is taken to be in a mixed state (e.g. for thermal averaging over the $j_g$). But even in this case the collective wall coordinate would remain to be purely mechanical without a temperature attached to it.

So, for this specific example as well as for specific choices of the parameters of the model (such that the perturbation calculation works), the correlation between the two subsystems reproduces what is known from the classical world. Through what we call a mechanical embedding, we have been able to test and verify the correctness of the widespread (e.g.~\cite{Bender}\cite{Bender2}) usage of the term 'pressure' in quantum systems. 


\textbf{Variance} - In order to clarify the quantum features of our model, the variance
\begin{equation}
\label{eq_variance}
(\Delta x_{\text{W}})^2=\erwwert{x_{\text{W}}^2}-\erwwert{x_{\text{W}}}^2
\end{equation}
of the position of the wall will be calculated. $\erwwert{x_{\text{W}}}$ is taken from eq.~(\ref{gl_QM_erwwert_x_W_ergebnis}), while $\erwwert{x_{\text{W}}^2}$ has to be derived analogous to eq.~(\ref{eq_erwwert_x}):
\begin{equation}
\erwwert{x_{\text{W}}^2} = {}'\bra{j_g,j_{\text{W}}}x_{\text{W}}^2\ket{j_g,j_{\text{W}}}'\approx {}^{\;'}_{\text{W}}\bra{j_{\text{W}}}\left( \frac{y_{\text{W}}^3}{L}+y_{\text{W}}^2\right)\ket{j_{\text{W}}}'_{\text{W}}
\end{equation}
where the $\ket{j_{\text{W}}}'_{\text{W}}$ are the 'wall-part' of the perturbed state from eq. (\ref{gl_QM_Störungsrechnung}). After some rearrangements one arrives at
\begin{equation}
(\Delta x_{\text{W}})_{j_g}^2\approx \frac{3\pi^2j_g^2}{2}\frac{\lambda^6}{\beta^2}L^2+\frac{1}{2}L^2\lambda^2-\frac{3}{4}\lambda^4L^2-\pi^4j_g^4\frac{\lambda^8}{\beta^4}L^2.
\end{equation}
The variance of the position of the wall has therefore the same order in the parameters $\{\lambda,\beta\}$ as the expectation value itself ($\erwwert{x_{\text{W}}}_{j_g} = \pi^2j_g^2L\frac{\lambda^4}{\beta^2}$). This shows the still dominating quantum nature of our model. In the classical limit the variance would be expected to decrease in comparison with the expectation value $\erwwert{x_{\text{W}}}$.


\textbf{Entanglement} - The interaction between the two subsystems 'gas' and 'wall' should lead to entanglement: The total system cannot be described by a single product state any more but rather as a sum over such states (eq.~(\ref{gl_QM_Störungsrechnung})). A measure for the amount of entanglement is the purity $P$ of one subsystem, e.g. the wall, which is the trace of the square of the reduced density operator $\hat{\rho}^{\text{W}}$ of the wall:
\begin{equation}
P^{\text{W}}=\text{Tr}\left\{ \left( \hat{\rho}^{\text{W}}\right)^2\right\}\,.
\end{equation}
The reduced density operator is obtained by 'tracing out' the other subsystem, the gas, of the density operator $\hat{\rho}$ of the pure total system:
\begin{equation}
\hat{\rho}=\frac{1}{N^2}\ket{j_g,j_{\text{W}}}'{}'\bra{j_g,j_{\text{W}}}\,; \quad
\hat{\rho}^{\text{W}}=\sum_l{}^{0}_{g}\bra{l}\hat{\rho}\ket{l}^0_{g}\,.
\end{equation}
$N$ is the normalization factor (eq.~(\ref{eq_normalization_factor})).

With the perturbed state from eq.~(\ref{gl_QM_Störungsrechnung}) one arrives at
\begin{equation}
P^{\text{W},0}\approx \frac{1}{N^4}\left[1+\pi^4 \frac{\lambda^6}{\beta^4}-\pi^2\frac{\lambda^4}{\beta^2}+\frac{1}{8}\lambda^2\right]
\end{equation}
for the purity of the subsystem wall when the total system is in its ground state ($j_g=1, j_{\text{W}}=0$). For a typical value $\lambda=\beta=10^{-3}$ for which this perturbation theory is valid, the subsystem purity is
\begin{equation}
P^{\text{W},0}\approx 1-8\cdot 10^{-9}
\end{equation}
and thus close to maximum. Correspondingly, the reduced von Neumann entropy will be very small. As should have been expected, the mechanical embedding does not introduce significant entropy.


\textbf{Generalization to three dimensions} - To get closer to thermodynamics, where one typically works with 'pressure' rather than with 'force', the results obtained so far should be generalized to a three dimensional model. This can be done easily because the two additional degrees of freedom do not depend on the position of the wall.
If the dimensions of the box in the three orthogonal directions are $L_1, L_2$ and $L_3$ and the corresponding quantum numbers are $j_{g,1}, j_{g,2}$ and $j_{g,3}$, the total energy of the particle in this box can be written using the volume $V=L_1L_2L_3$ as
\begin{equation}
E^{g} = \frac{\pi^2\hbar^2}{2M_g}\left( \frac{j_{g,1}^2}{L_1^2}+\frac{j_{g,2}^2}{L_2^2}+\frac{j_{g,3}^2}{L_3^2}\right) = \frac{\pi^2\hbar^2}{2M_g}\left( \frac{j_{g,1}^2L_2^2L_3^2}{V^2}+\frac{j_{g,2}^2}{L_2^2}+\frac{j_{g,3}^2}{L_3^2}\right)\,.
\end{equation}
The 'classical' result for the pressure of this 1-particle-gas is
\begin{equation}
\label{gl_QM_3D_klass}
p=-\frac{\partial E^{g}}{\partial V}=\frac{\pi^2\hbar^2}{M_g}\frac{j_{g,1}^2L_2^2L_3^2}{V^3}\,.
\end{equation}

Because the two new degrees of freedom do not affect the results of the perturbation theory, the result for the expectation value of the position of the wall is the same (eq.~(\ref{gl_QM_erwwert_x_W_ergebnis})). If one puts in $L=\frac{V}{L_2L_3}$ in eq.~(\ref{gl_QM_erwwert_x_W_ergebnis}) and uses the fact that pressure is force divided by area ($L_2L_3$), the equivalence between the classical and the quantum mechanical result is the same as in the one dimensional case.

\section{Summary}
In this paper the dependence of a spatial coordinate of a two-particle quantum mechanical system on the energy of one of the subsystems has been studied. The spatial coordinate considered - the expectation value of the position of a harmonic oscillator (representing a reacting wall) - can be taken as an external parameter of the system, and we were able to show that the classical thermodynamic result for the pressure, $p=-\pop{E}{V}$, has an analogy in the quantum world for specific values of the system parameters. Under these conditions it is justified even in the quantum regime to call $-\pop{E}{V}$ a 'pressure'. However, this is a stationary property of a quantum system embedded in a mechanical environment and does not support the classical picture of ``thermal motion'' (i.e. a particle bouncing off the wall).

The quantum character of our system manifests itself in its variance which is found to be of the same order of magnitude as the expectation value of $x_{\text{W}}$, as is typical for quantum states in a non-classical regime. For higher quantum numbers $j_g$ in the gas $g$ (i.e. higher energies) the relation between expectation value and variance decreases as expected for the transition to classical systems. We expect the same trend for a quantum-manometer with more than one particle, but have not carried out the corresponding calculations since the present perturbation theory cannot handle the case of degenerate energy levels present in such larger systems.

The entanglement between the two subsystems, though shown to be very small for typical cases, is nevertheless generic for having the particle interact with the wall.

Our closed bi-partite system may be considered part of a complete measurement scenario (if supplemented, e.g., by continuous observation of the position of the piston).

\acknowledgments
Financial support by the Deutsche Forschungsgemeinschaft is gratefully acknowledged.

%

\end{document}